# In Pursuit of 2D Materials for Maximum Optical Response

Sunny Gupta, Sharmila N. Shirodkar, Alex Kutana and Boris I. Yakobson[*]
*Department of Materials Science and NanoEngineering, Rice University, Houston, TX, 77005*

**ABSTRACT**
Despite being only a few atoms thick, single-layer two-dimensional (2D) materials display strong electron-photon interactions that could be utilized in efficient light modulators on extreme subwavelength scales. In various applications involving light modulation and manipulation, materials with strong optical response at different wavelengths are required. Using qualitative analytical modeling and first-principles calculations, we determine the theoretical limit of the maximum optical response such as absorbance ($A$) and reflectance ($R$) in 2D materials and also conduct a computational survey to seek out those with best $A$ and $R$ in various frequency ranges, from mid-infrared (IR) to deep ultraviolet (UV). We find that 2D boron has broadband reflectance $R$ >99% for >100 layers, surpassing conventional thin films of bulk metals such as silver. Moreover, we identify 2D monolayer semiconductors with maximum response, for which we obtain quantitative estimates by calculating quasiparticle energies and accounting for excitonic effects by solving the Bethe-Salpeter equation (BSE). We found several monolayer semiconductors with absorbances ≳30% in different optical ranges which are more than half of the maximum possible value for a 2D material. Our study predicts 2D materials which can potentially be used in ultra-thin reflectors and absorbers for optoelectronic application in various frequency ranges.

Due to their reduced dimensionality, two-dimensional (2D) materials exhibit an extraordinary optical response in comparison with bulk counterparts,[1–4] as has been shown early on with the examples of graphene and 2D MoS$_2$.[5–7] The most well-known of 2D materials, graphene, has its low frequency absorbance defined by the universal analytical constant ($\pi/137$) ~ 0.023.[1] In 2D, the joint density of states can exhibit logarithmic singularities,[8] resulting in enhanced absorption. In addition, spatial confinement and reduced dielectric screening of 2D materials causes strong Coulomb interactions that beget more stable exciton formation with large binding energy and oscillator strength compared to bulk crystals,[9] also enhancing their optical properties. Being a prototypical 2D semiconductor, 2D MoS$_2$ has exceptional optical absorption/photoluminescence in the visible range. Its astounding optical properties have opened up prospects for 2D materials exploration for use as absorbers, reflectors, and modulators[10] in optical nanodevices such as photodiodes, solar cells, photocatalytic cells, phototransistors, and photodetectors.[11–16] These materials host stable room-temperature excitons, and are ideal candidates for understanding light-matter interactions and possible application in development of excitonic polariton devices.[17]

For a variety of applications, 2D materials with strong optical response (such as absorbance and reflectance) in different frequency ranges (mid-, near-infrared (IR), visible, near-, mid-, deep-ultraviolet (UV)) are necessary, but a cohesive study estimating the transmittance, absorbance, and reflectance (*TAR*) of 2D materials is still lacking. Also, maximum achievable values of *TAR* in 2D materials have not been analyzed. For instance, it is unclear whether there are any limitations on *TAR* in 2D, *e.g.* whether a 2D layer could be as absorbing/reflecting as a bulk material. Being at the ultimate limit of atomic size in one direction, 2D materials can also serve as elementary building blocks in more sophisticated structures such as metamaterials, where a nontrivial layer-dependent optical response emerges. It is of interest to find out atomically thin monolayers that can provide the strongest absorbance/reflectance and how close these can approach the limiting values for a 2D material.

Here, we use first-principles calculations to evaluate the *TAR* of a wide variety of 2D materials (55 monolayers), over a wide optical spectrum, to identify and quantify the materials with strongest response. Among them, most synthesized and predicted 2D materials for optical applications are semiconductors while intrinsic graphene is a semimetal. Recently, 2D polymorphs of boron (borophene), have attracted great interest,[18,19] after reports of successful synthesis.[20,21] They are intrinsically metallic with much higher numbers of free carriers than doped graphene or semiconductors. While there is rich polymorphism in this material, [22] here we restrict ourselves to the triangular polymorph with *p2mm* symmetry,[23] and calculate its in-plane conductivity along (*x*) and normal to (*y*) its zig-zag buckled direction. We found borophene heterostructures to show broadband reflectance with $R$ >99% for >100 layers from IR to UV range, which is superior to bulk metals such as even silver. Moreover, in 2D semiconductors, band nesting and excitonic effects result in high absorbance, and we screen out best absorbers having these properties in each region of the optical spectrum. Our findings reveal and quantify basic properties across a 2D material family as well as identify materials holding promise for design of ultracompact optoelectronics in a wide frequency range.

**RESULTS AND DISCUSSION**

A 2D material can be viewed as a zero-thickness layer between two semi-infinite dielectric media-slabs,[24–26] and its *TAR* can be obtained using the transfer matrix formulation for waves in layered piecewise-constant media applied to a single interface.[25,27–29] In case of a 2D interface, the transfer matrix connects the amplitudes of the normally-incident and reflected waves across the interface and yields the following (for details see S1, Supporting Information),

$$T = \frac{4n_1 n_2}{|n_1+n_2+\sigma_{2D}Z_{vac}|^2} \quad (1a)$$

$$A = \frac{4n_1 Re\{\sigma_{2D}\}Z_{vac}}{|n_1+n_2+\sigma_{2D}Z_{vac}|^2} \quad (1b)$$

$$R = \left|\frac{n_2-n_1+\sigma_{2D}Z_{vac}}{n_1+n_2+\sigma_{2D}Z_{vac}}\right|^2 \tag{1c}$$

Here $n_{1,2}$ are the refractive indexes on either side of monolayer, $\sigma_{2D} = \sigma_{2D}(\mathbf{q},\omega)$ is the optical surface conductivity as a function of frequency ($\omega$) and 2D wavevector ($\mathbf{q}$), $Z_{vac} = 376.73\ \Omega = 1/\varepsilon_0 c$ is the impedance of vacuum. Under normal incidence condition, the in-plane component of $\mathbf{q}$ is zero; we henceforth restrict our discussion only to this case. The conductivity of a material determines its optical properties and is estimated from the linear dielectric response as:

$$\sigma_{2D}(\mathbf{q}=0,\omega) = i\omega\varepsilon_0(1-\varepsilon_{3D}(\mathbf{q}=0,\omega))L. \tag{2}$$

where $\varepsilon_{3D}(\mathbf{q},\omega)$ is the head of the full dielectric function calculated for the 2D layer separated from its periodic images by vacuum $L$. We assume the $\mathbf{q}=0$ limit and drop the symbol in expressions henceforth.

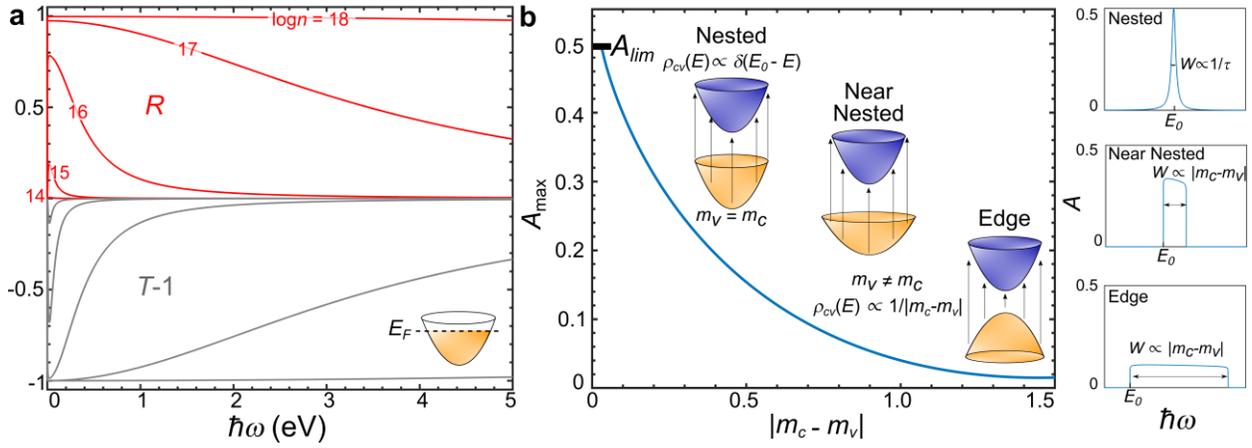

**Figure 1.** (a) Reflectance $R$ and transmittance $T$ of a model two-dimensional metal described as a 2D electron gas for electron concentrations $n$ between $10^{14}$ and $10^{18}$ cm$^{-2}$. The values for absorbance $A$ in the visible range are small (<0.01) and not shown for clarity. Other fixed parameters are $m = m_e$, $\tau = 30$ fs, and $k_F^2 = 2\pi n$. (b) Calculated maximum absorbance of a model two-band semiconductor with different joint densities of states ($\rho_{cv}$) depending on the effective masses of the valence ($m_v$) and conduction bands ($m_c$) as a function of $|m_c–m_v|$. Perfect nesting, $m_c = m_v$; absorbance has a narrow peak with the width inversely proportional to scattering time $\tau$ and maximum value reaching $A_{lim} = 1/2$ (upper right panel). Nearly perfect nesting, $m_c \neq m_v$ and $m_c, m_v > 0$; $A < 1/2$, and the width of the absorbance peak is proportional to $|m_c–m_v|$ (middle right panel). Absorption edge, $m_c \neq m_v$, $m_c > 0$, $m_v < 0$, yielding a step-like feature in absorbance (lower right panel).

Phenomenological Eq. 1 already reveals a few obviously important features. First, the $TAR$ sum is invariant, $T + A + R = 1$, correctly capturing energy conservation. Second, at very high conductivity, $\sigma_{2D}(\omega) \to \infty$, the layer can reflect fully, $R \approx 1$. (The condition for achieving maximum $R = 1$ in 2D materials from Eq. 1c is different from that in 3D, where $R = 1$ for any negative real values of dielectric constant.) Third, it is straightforward to see that maximum of Eq. 1b is $A_{max} = 0.5$ (at $\sigma' = 2$ and $\sigma'' = 0$, with $\sigma_{2D}Z_{vac} \equiv \sigma' + i\cdot\sigma''$) that is absorption reaches 50%, while $T = R = 25\%$. Further microscopic insight can be gleaned from simple physical models as follows.

The first one is 2DEG, a 2D electron gas, representing a metal layer in vacuum, ($n_1 = n_2 = 1$ in Eq. 1). Its 2D conductivity based on Lindhard dielectric function within the random phase approximation (RPA) can be obtained analytically (see S2, Supporting Information), and yields

$$\sigma_{2D}(\omega) = \frac{i\omega}{m} \frac{ne^2}{(\omega+i/\tau)^2} \qquad (3)$$

Here, $n$ is the 2D carrier density, $m$ is the effective mass of electron, $\tau$ is the phenomenological scattering time. The contribution to $\tau$ is from electron-electron, electron-phonon and radiative scattering mechanisms. Since we consider pristine, flat and free standing 2D material, the scattering due to impurities, defects and surface roughness is neglected. The *TAR* are evaluated by substituting Eq. 3 into Eq. 1 and plotted in Figure 1a for different $n$, keeping other parameters at their generic values: $m = m_e$ as for free electron and $\tau = 30$ fs. As a single-band metal, such 2DEG has very low $A$, while its $R$ increases with $n$ (Figure 1a). It follows from Eq. 1c and 3 that for the long waves like THz-radiation with $\omega\tau \sim 1$, if the carrier concentration is large in proportion to the relaxation rate, $n \gg \varepsilon_0 e^{-2} mc/\tau$, the 2DEG can be a good reflector with $R \approx 1$. However, in the more common visible range at $\omega\tau \gg 1$, the requirement for strong reflection is rather stringent, $n \gg \varepsilon_0 e^{-2} mc/\omega$ (see S3, Supporting Information); an estimated $n \sim 10^{17}$-$10^{18}$ cm$^{-2}$ appears unreasonably high for atomic monolayer but perhaps achievable as a sum over some thickness. Indeed, later we will see that $R$ increases by stacking 2D materials in heterostructure geometry, effectively corresponding to large $n$.

More instructive in this respect is the second model we examine, a *multiband* case with interband transitions. It can be viewed as electron gas confined by a $\delta z$-narrow potential well, in a direction normal to its free $x$-, $y$-coordinates; if $\delta z \to 0$, the quantum confinement raises energy spacing so much that only single band remains. It is reduced to the two-band (in other words, two-level system with dispersion) case when the transition energy is close to the distance between two bands among many. Using Fermi golden rule, the imaginary part of the dielectric constant $\varepsilon_{2D,imag}$ in a multiband system is given by Eq. S4.4 (Supporting Information), which is proportional to $|\langle v|p|c\rangle|^2$, the square of the matrix element of momentum, and the joint density of states (JDOS) $\rho_{cv}(\omega) = \sum_k \delta(E_c(k) - E_v(k) - \hbar\omega)$. Introducing phenomenological scattering time $\tau$, we cast $\varepsilon_{2D}$ into $\sigma_{2D}$ in the Lorentz model as:

$$\sigma_{2D}(\omega) = \frac{i\omega}{m} \sum_j \frac{f_j ne^2}{(\omega^2 - \omega^2_{0j}) + i\omega/\tau}, \qquad \sum_j f_j = 1 \qquad (4)$$

Here, $\omega_{0j}$ is the frequency of $j^{th}$ transition, and $f_j$ is the oscillator strength. From Eq. S4.4, $f_j = \frac{2|\langle v|p|c\rangle|^2}{m\hbar\omega_{0j}}$, expressing the conservation of the total number of electrons. The case $\omega_{0j} = 0$ corresponds to intraband transitions and finite frequencies represent interband transitions. We next consider different cases for JDOS depending on the effective masses of the valence ($m_v$) and conduction bands ($m_c$) and use it to calculate *TAR* under different conditions (for details see S4, Supporting Information). The left panel of Figure 1b shows the calculated maximum absorbance $A_{max}$ as a function of the difference of the effective masses in the conduction and valence bands, $|m_c - m_v|$.

For *perfect* nesting, $m_c = m_v$. All electrons in the fully occupied band participate in the transition to the empty band at energy $E_0 = E_c - E_v =$ const. The upper right panel of Figure 1b shows the absorbance as a function of the photon energy. For perfectly nested bands, $\omega_{0j} = \omega_0 = E_0/\hbar$, and $\sigma_{2D,real} \propto \delta(\omega - \omega_0)$ with $f_j = f_0 = 1$. The width of the absorption peak is determined by the scattering time in the upper and lower bands, due to non-zero imaginary part of electron-phonon and electron-electron self-energy. The $\delta$-function is therefore replaced with a Lorentzian with width $1/\tau$, and the maximum absorbance $A_{\text{lim}}=1/2$ is achieved when $c/\tau_{A,\text{max}} = ne^2/2\varepsilon_0 m$, where $c$ is the speed of light. In case of perfect nesting, maximum absorbance can be achieved even when carrier concentration $n$ is low; however, the requirement for the carrier scattering time becomes more stringent as concentration decreases. When the condition for maximum absorbance is satisfied, one has $T = R = 1/4$, and for $\tau > \tau_{A,\text{max}}$, $R > T$, whereas for $\tau < \tau_{A,\text{max}}$, $T > R$.

For nearly perfect nesting, $m_c \neq m_v$, $\Delta E >> \hbar/\tau$, where $\Delta E$ is the energy window where nesting is significant. In this case there is a peak in absorption in a narrow range of frequencies around the nesting transition. The absorbance profile for this case is shown in the middle right panel of Figure 1b. Unlike the case of perfect nesting, the width of the peak is determined by $\Delta E$ instead of $\tau$. For parabolic bands considered here, $\Delta E = \hbar^2 k_c^2 |m_c - m_v|/2m_c m_v$, where $k_c$ is the cutoff wavevector beyond which band nesting becomes weak. The JDOS in this case is constant and proportional to $m_c m_v / |m_c - m_v|$, and absorbance $A < 1/2$. The maximum absorbance $A$ increases with decreasing $|m_c - m_v|$, approaching the limit of $A_{\text{lim}} = 1/2$ when $\Delta E \sim \hbar/\tau$, as seen in the left panel of Figure 1b. This underscores the importance of existence of a high degree of band nesting in a material to achieve high absorbance.

For absorption edge, $m_c \neq m_v$, $m_c > 0$, $m_v < 0$. While formally this case is similar to case of nearly perfect nesting, it is considered separately here in order to emphasize the absence of a sharp absorption peak due to large difference in effective masses. In two dimensions, this corresponds to the JDOS being constant above transition frequency $\omega_0 = E_0/\hbar$ i.e. continuous interband transitions up to $E_0 + \Delta E$, $\Delta E = \hbar^2 k_c^2 |m_c - m_v|/2m_c m_v$. Similar to case of nearly perfect nesting, maximum absorbance $A_{\text{max}}$ can approach the value $A_{\text{lim}} = 1/2$ when $|m_c - m_v|$ is small.

Bearing in mind the analytical results for maximum limits on the $T$, $A$, and $R$ for 2DEG, we now turn to DFT calculations for determining the optical properties of 2D materials. We find that DFT yields the Kohn-Sham band gap of ~1.8 eV in MoS$_2$, which at that level of theory is close to the threshold frequency of absorption, whereas the quasiparticle gap obtained with $G_0W_0$ method is ~2.41 eV.[5,30–33] On the inclusion of electron-hole corrections an excitonic peak appears at ~1.9 eV, reducing the absorption threshold closer to the DFT band gap estimate (see S3 and Figure S5, Supporting Information for details). Thus, although DFT cannot capture excitonic effects, its threshold frequency and the frequency of maximum absorbance are accurately determined within this approximation, due to cancellations of various corrections, which is in agreement with previous studies on MoS$_2$.[5,30–32] We find this trend to be true for other materials, such as H-MoTe$_2$, T-PtTe$_2$, H-TiS$_2$, and T-SnSe$_2$. Hence, our initial screening of *TAR* is performed at the DFT level, without $G_0W_0$+BSE corrections. (see S6, Supporting Information for results of unconverged $G_0W_0$+BSE corrections).

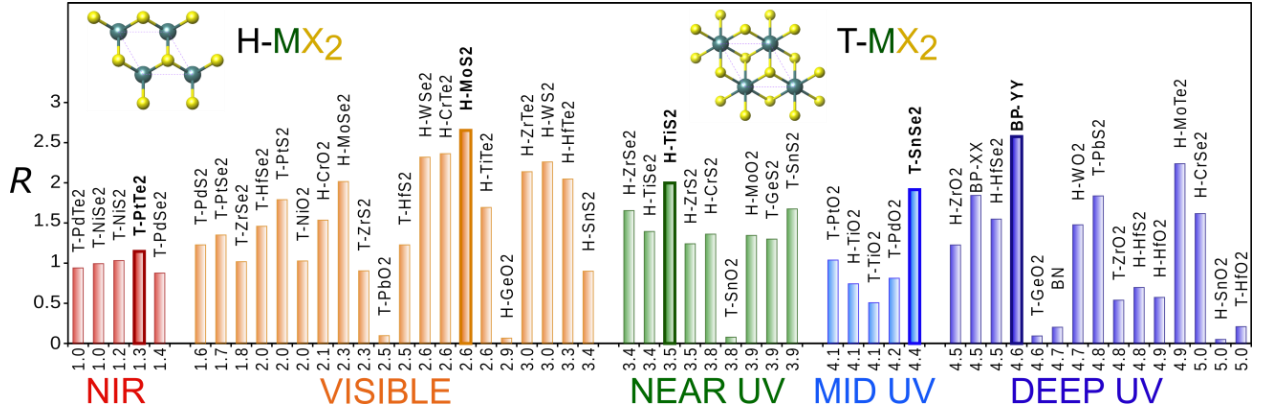

**Figure 2**. Reflectance, $R$ (%) of 53 monolayers. The best reflective material in each region is highlighted.

The dielectric function at DFT level (under RPA) was used to estimate $\sigma_{2D}$ (Eq. 2) and *TAR* (Eq. 1). The calculated *TAR* for freestanding monolayer graphene and the triangular polymorph of two-dimensional (2D) boron, is found to be in good agreement with existing literature (Figure S6, Supporting Information).[34,35] Moreover, the estimate of $A(\omega \to 0) = 0.023$ for graphene is close to the universal analytical constant $(\pi/137)$,[1] showing that the used methodology is accurate. The *TAR* are a strong function of $\omega$, as evident from the above examples of graphene and 2D boron, and thus comparison of different 2D materials should be done at specific frequency intervals. Here we restrict our analysis to energies below 5 eV, since most of the applications involving photodetection do not go beyond the deep UV range. We now focus on the reflectance of 2D semiconducting monolayers. The maximum of $R$ for $\hbar\omega < 5$ eV, and corresponding frequencies for the 53 2D semiconductors is plotted in Figure 2. Recently a large number of them were experimentally synthesized and a generalized procedure to make the others was proposed.[36] The materials were assumed to be freestanding, *i.e.* adjacent to vacuum on both sides. The optical properties strongly depend on the relaxation time $\tau$, which is determined by several scattering mechanisms: electron-electron (e-e), electron-phonon (e-ph), impurity, defect, surface roughness scattering as well as natural linewidth (n). In our idealized model of pristine flat and free standing 2D layers, the impurity, defect and surface scattering mechanisms give null contribution. Hence the total $1/\tau = 1/\tau_{e-e} + 1/\tau_{e-ph} + 1/\tau_n$, and is dominated by the shortest scattering time. Recent works report, $\tau_{e-e} \sim \tau_{e-ph} \sim 13$ fs[37,38] in MoS$_2$ and 22 fs[39] in MoSe$_2$ respectively. At the same time, $\tau_n$ is at least 100 times longer in these materials,[39,40] and hence $\tau \sim \tau_{e-e}, \tau_{e-ph}$. In view of the scarcity of values in literature and extensive computational costs for theoretical estimates of $\tau$, we empirically apply the values similar to ones reported for MoSe$_2$ and MoS$_2$ to all materials. Hence, we choose a reasonable $\tau \sim 13$ fs for all 53 semiconductors in our work. We find that T-PtTe$_2$, H-MoS$_2$, H-TiS$_2$, T-SnSe$_2$, and black phosphorus (BP along *y* direction; BP-YY) are the most reflecting materials in different optical regimes as shown in Figure 2. We note that since these materials are semiconductors, their response is solely due to interband transitions resulting in appreciable absorbance that lowers the reflectance.

Although the $R$ of free-standing monolayers is small, it has been shown by Papadakis et al.[41] that heterostructures constructed from stacks of graphene and hexagonal boron nitride show 99.7% reflectance in the mid-IR ranges at a fraction of the weight of noble metals. Based on this approach, we extend our work to study the effect of stacking on the $TAR$ properties of these materials in the visible frequency range (~2 eV) using the transfer matrix method.[26,27] We benchmark our calculations using graphene (doped to $E_F$ ~ 0.2 eV and with relaxation time of ~300 fs) as a reference to compare with previous literature,[41] and find our results for $R$ of 250 layers with spacing of 0.67 nm to agree within ± 0.5% at $\lambda$ = 40 μm. Slight discrepancy may arise from the difference in other parameters such as Brillouin Zone (BZ) sampling. In the current work, a 600×600×1 $k$-point mesh was used. Since the graphene electronic structure and optical response are symmetric with respect to $p$- and $n$-doping at low concentrations, only one type of doping needs to be considered. Here, we consider $n$-doped graphene.

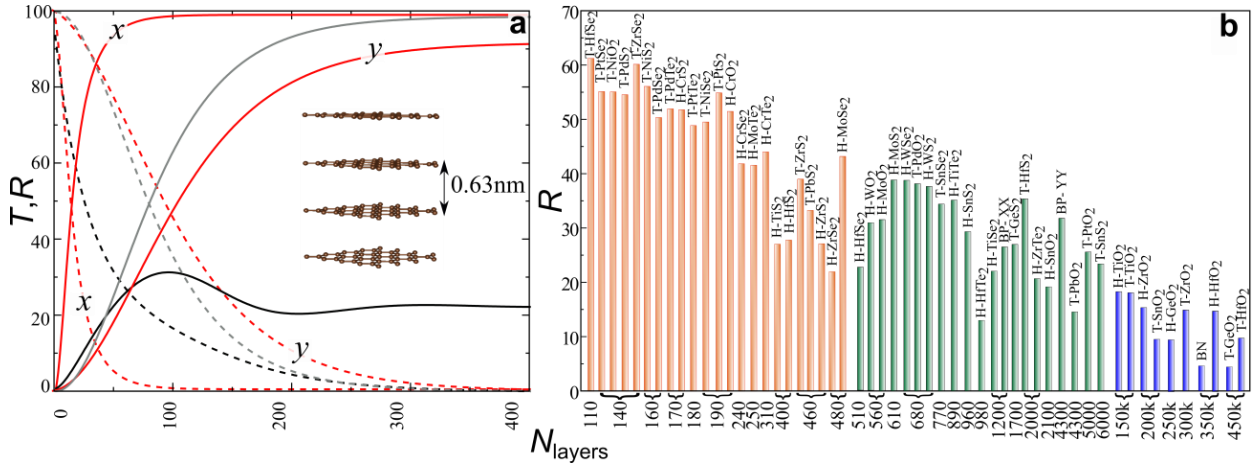

**Figure 3.** Evolution of (a) $T$ (%) (dashed line) and $R$ (%) (solid line) with number of layers of triangular 2D boron (red) for $x$ and $y$ polarization, doped graphene ($E_F$ = 0.2 eV, black), and bulk silver film (gray) of corresponding thickness using Lindhard model. $T$ decays exponentially with number of layers. (b) Minimum number of layers required for saturated $R$ (%) (with $T = 10^{-3}$%) of 53 semiconductor monolayer stacks. All results are at 620 nm (2 eV) of incident light with spacing 0.63 nm between layers.

Unlike graphene, 2D boron is intrinsically metallic with equivalent $n = 5.5 \times 10^{15}$ cm$^{-2}$, making it an effective single-band metal for frequencies including visible range. Also, its structural anisotropy is reflected in the anisotropy of the $TAR$ coefficients for different polarizations of light. Due to the high metallicity and anisotropic $TAR$ coefficients, 2D boron exhibits quite high $A$ and $R$ as compared to graphene (see S7, Supporting Information for details), hence outperforming it in the visible-range. However, the performance of 2D boron with respect to the number of layers has not yet been considered to the best of our knowledge. Moreover, 2D boron exfoliation from substrate into a free layer has not been achieved yet. In Figure 3a, we compare how the response of 2D boron and doped graphene ($E_F$ = 0.2 eV) changes with number of layers at the visible wavelength, $\lambda$ ~ 620 nm ($\hbar\omega$ = 2 eV), at interlayer separation of 0.63 nm. Our choice of spacing is such that the interlayer interaction is weak, and layers can be treated as independent when applying the transfer matrix method. The calculations for 2D boron were carried out on a 300×300×1 $k$-point mesh with a relaxation time of ~29 fs characteristic of bulk metals.[42] We find that $R$ in 2D boron for $x$ polarized light reaches ~98.9% for >100 layers, whereas $R$ of doped graphene saturates at ~21.8% for >200 layers. For $y$ polarization, reflectance of 2D boron is not as high, reaching ~90% for >300 layers. More importantly, the $A$ in multilayer 2D boron is ~0%, as compared with $A \sim 78.1\%$ in multilayer doped graphene. This high reflectance and low absorption in 2D boron is due to absence of interband transitions up to 3.4 eV, whereas in graphene the onset of interband transitions is at $2E_F = 0.4$ eV, making highly metallic 2D boron far superior to doped graphene as a reflector in visible range. Also, since graphene needs external (chemical or gate) doping to make it metallic, which could potentially introduce losses or change its electronic structure; whereas 2D boron is intrinsically metallic thus making it a more attractive reflector material. The advantage of the high threshold for the onset of interband transitions is also clear when comparing $R$ of boron with films of bulk transition metals. In Figure 3a we show $R$ and $T$ of a silver thin film represented by the Lindhard model with plasma frequency 3.8 eV and relaxation time of 31 fs.[42] The 2D boron heterostructure has better reflectance than metal film of the same thickness. Moreover, this heterostructure also shows broadband reflectance from IR to UV range (Figure S7, Supporting Information), ideal for designing ultrathin reflectors in different optical region. These heterostructures could possibly be used for designing coating materials for protection from high intensity lasers. The 2D boron as well as graphene and hexagonal boron-nitride heterostructures[41] outperform metal thin films at long wavelengths, suggesting potential advantage of 2D heterostructures over bulk materials.

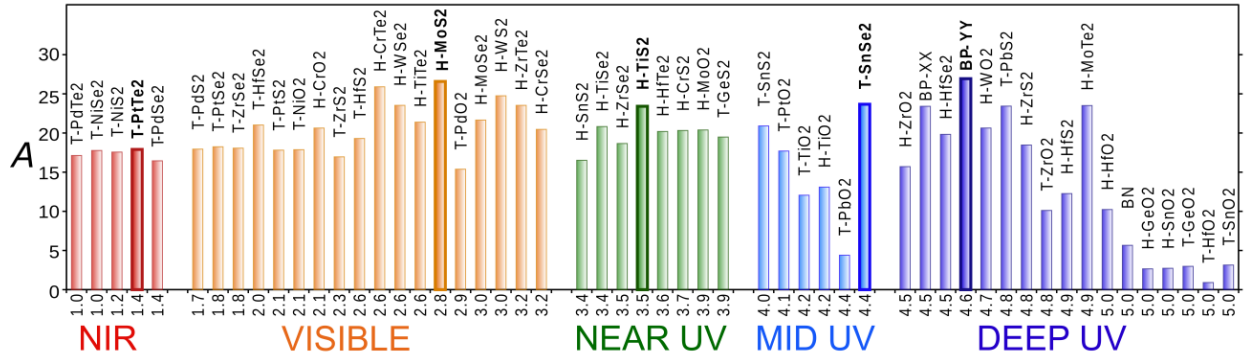

**Figure 4**. Absorbance, $A$ (%) of 53 monolayers. The best absorbing material in each region is highlighted.

We extend the analysis to the other 53 semiconductors and estimate the minimum number of layers required for $R$ to reach constant value (with zero transmission) at $\lambda \sim 620$ nm, as shown in Figure 3b. Similar to the case of 2D boron heterostructures, the distance between the interlayer chalcogens in semiconductor layers was chosen as 0.63 nm (approximately twice the normal equilibrium van der Waals distance) so that the layers are decoupled and transfer matrix description can be applied. Hereafter, this value of interlayer distance is used for all heterostructures. This larger interlayer separation can be practically achieved by inserting inert materials such as 2D BN between the layers. To a lesser degree, the interlayer coupling may be decreased by mutually rotating the layers to incoherent turbostratic stacking. It is important to note that, when thickness of the material increases, the phase difference of transmitting light between the layers (which is dependent on $\lambda$) increases and the interference effects from multiple reflections across layers become significant, leading to oscillations in the *TAR*, as *e.g.* seen as a peak in *R* for graphene in Figure 3a. Here we report the number of layers needed to reach saturation in *R* (with $T = 10^{-3}$%). The general trend in Figure 3b shows that materials that reach saturation faster yield higher *R* values, with a maximum $R \sim 62$% for T-HfSe$_2$ at 110 layers. Reflectance for T-PtTe$_2$, H-MoS$_2$, H-TiS$_2$, T-SnSe$_2$ and BP along *y* direction saturates at ~48% (180 layers), 39% (610 layers), 27% (400 layers), 34% (770 layers) and 33% (4300 layers), respectively. Other materials, which show saturation at few thousands of layers, are clearly better absorbers/transmitters than reflectors as thin films. Our analysis shows that metals, especially 2D boron is a good reflector even at visible frequencies, as opposed to semiconductors and doped graphene which are better absorbers at these wavelengths. These structures open up possibilities for more complex optical metamaterials from 2D layers, as their characteristic dimension *a* satisfies the condition $\lambda/a \sim 100$ for the metamaterials,[27] yielding $a \sim 5$ nm at a typical optical wavelength $\lambda \sim 500$ nm.

Although 2D semiconductors are not expected to show good reflectance under ambient conditions, they have been found to possess good absorbance. We next study the absorbance of monolayers of 53 2D semiconductors. The maximum of *A* for $\hbar\omega < 5$ eV, and their corresponding frequencies for all materials are plotted in Figure 4. The maximum response is solely due to interband transitions. We screen for materials with best absorbance in each frequency region and find that the best performing materials are T-PtTe$_2$, H-MoS$_2$, H-TiS$_2$, T-SnSe$_2$ and BP along *y* direction, in different regions of the electromagnetic spectrum respectively, and BP-YY being the highest absorbing material with $A \sim 27$% as shown in Figure 4. Ranking materials according to maximum *R, A* in a narrow frequency range may not be suitable for all applications. A large (if not maximum) *R*, *A* in a broader frequency range may be preferable in some cases, wherein the integral of *R*, *A* with respect to frequency is maximized. For this purpose, different 2D materials with maximum *R*, *A* in different optical regimes can be vertically stacked into heterostructures to achieve broadband response, thus further expanding the scope of 2D materials in optical nanodevices, as further discussed below.

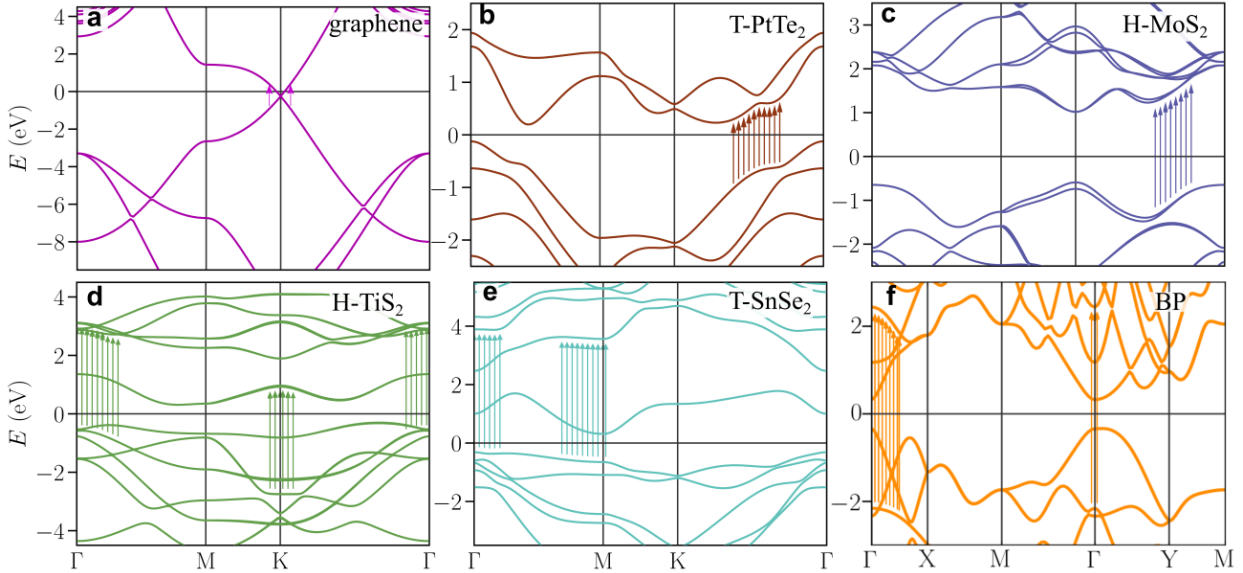

**Figure 5**. Band structures of (a) *n*-doped graphene, (b) T-PtTe$_2$, (c) H-MoS$_2$, (d) H-TiS$_2$, (e) T-SnSe$_2$, and (f) BP along y direction. Vertical arrows show band nesting regions responsible for high absorbance.

To get further insights into the origins of the large optical response of the semiconducting monolayers in the different regions of the electromagnetic spectrum, we performed noncollinear band structure calculations of the selected 2D materials. The band structures of the materials with largest response are shown in Figure 5. They include metallic *n*-doped graphene, three indirect-gap semiconductors (T-PtTe$_2$, H-TiS$_2$, and T-SnSe$_2$), and two direct-gap semiconductors (H-MoS$_2$ and black phosphorus). The bands in H-type structures show small spin-orbit splitting, whereas splitting in T-type structures, black phosphorus and graphene is absent due to their spatial inversion symmetry. Vertical arrows show band nesting regions responsible for the absorption peaks at energies in each of the ranges indicated in Figure 4. Note that in all semiconductors, the transitions responsible for strongest absorption do not occur between band edges. This is expected, since bands have opposite curvatures at the edges, precluding existence of large sections of parallel bands, hence start of absorption corresponds to the case of absorption edge (Figure 1b). Such constraints are absent in general, and in comparison, to analytical results for 2DEG discussed in the previous section, nesting scenarios correspond to the case of nearly perfect nesting.

**Table 1.** *GW* and **DFT-PBE band gaps ($E_g$) of 2D semiconductors with strong optical response.**

| material | $E_g$ (eV) | |
| --- | --- | --- |
| | DFT-PBE | GW |
| T-PtTe$_2$ | 0.40 | 0.69 |
| BP | 0.90 | 1.80 |
| H-TiS$_2$ | 0.71 | 1.66 |
| T-SnSe$_2$ | 0.79 | 1.78 |

| H-MoS$_2$ | 1.62 | 2.41 |

In T-PtTe$_2$ and H-MoS$_2$, band nesting occurs between the valence band (VB) and conduction band (CB) at same regions of the BZ, namely along the ΓK line. The marked transitions in MoS$_2$ correspond to its well-known C excitonic peak, with the energy ~2.7 eV. We note that the absorbance in a single MoS$_2$ layer at the C peak (~26%, about half of the maximum possible value) is much larger than that at the A peak (~10%), consistent with better nesting conditions away from the band edges. The absorption peak of BP for polarization along Y direction at 4.6 eV originates from transitions at the BZ center (see Figure 5f). Unlike H-MoS$_2$, T-PtTe$_2$ is an indirect-gap semiconductor; however, in T-PtTe$_2$, there are similar transitions at band nesting yielding the absorbance of ~17% at 1.4 eV. Note that H-TiS$_2$, and T-SnSe$_2$ have nearly dispersionless bands both below and above the Fermi level which are responsible for their high absorbances at 3.5, and 4.4 eV respectively.

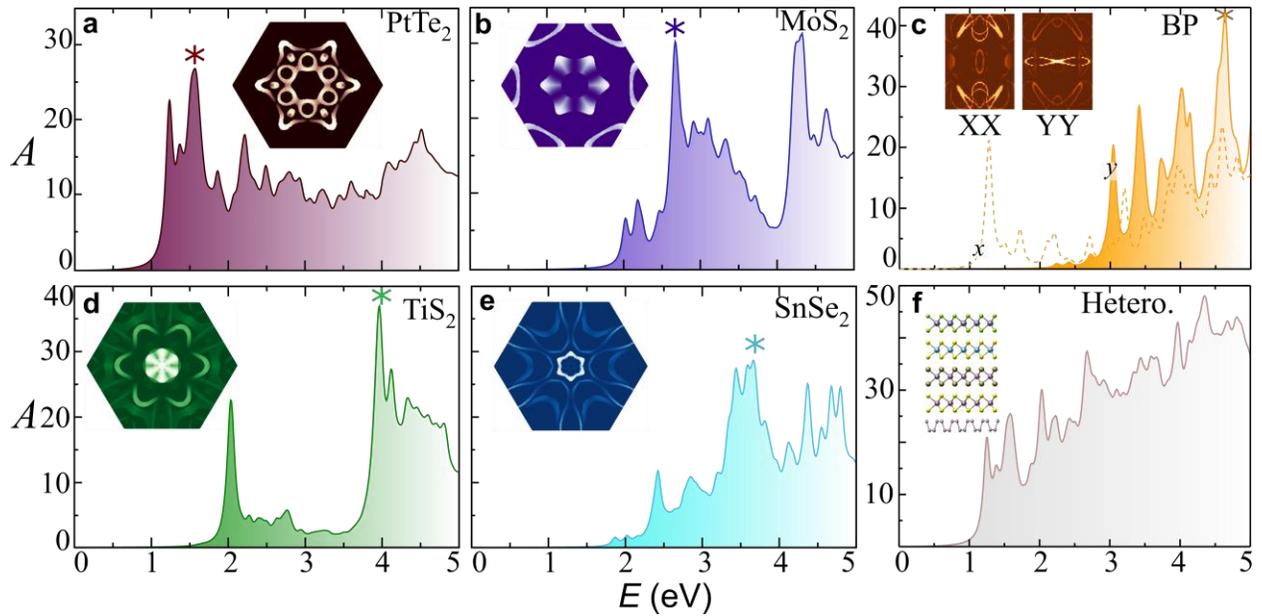

**Figure 6**. (a-e) BSE absorption spectra (%), with inset showing band nesting regions in full Brillouin Zone responsible for peaks denoted by ✶. (f) Absorption spectrum (%) of the heterostructure obtained by stacking all the 5 materials in a-e.

DFT can provide good qualitative insights into the essential features of the electronic structure, but to have quantitative predictions one needs to go beyond the one-particle picture of DFT. We next utilize quantitative treatments in the form of post-DFT many-body *GW* and BSE approaches. We performed calculations of the five 2D semiconductors at the $G_0W_0$+BSE level introducing self-energy corrections and capturing excitonic effects and ensuing optical properties. Table 1 shows the DFT-PBE and *GW* quasiparticle band gaps and the BSE absorption spectra are shown in Figure 6 (a-e). As expected, *GW* corrections to DFT gaps are significant in all cases, ranging from 0.3 to 1 eV. We estimated the dielectric function ($\varepsilon_{3D}(\omega)$) of monolayers in a supercell geometry with vacuum on both sides at the $G_0W_0$+BSE level. We then converted it to the in-plane conductivity (Eq. 2) and used Eq. 1 to estimate the *TAR*. A $\tau$ of ~13 fs same as in previous DFT calculations was used. We find that including excitonic effects indeed changes optical spectra substantially, in particular by introducing pronounced excitonic peaks and shifting peak positions, as compared to the one-particle DFT spectra. The BSE absorbance spectrum of T-PtTe$_2$ is shown in Figure 6a. We first note the absorbance peak at 1.5 eV, yielding the absorbance of 27% in a single monolayer, which is highest among all considered 2D TMDs in the near IR range. In case of H-MoS$_2$, the band nesting transitions yield an absorbance maximum at 2.7 eV, and the corresponding excitonic C peak is at ~2.66 eV, as seen in Figure 6b. Our BSE results for MoS$_2$ are in quantitative agreement with previous theoretical results and experiment;[5,30–33] in particular, the absorbance of ~30% at the C peak in H-MoS$_2$ compares favorably with the experimental value of ~25%[5,31] (a slight difference is due to the difference in $\tau$ between our calculations and experiments). In the absorption spectrum of BP-YY, shown in Figure 6c, we find record predicted absorbance of ~43% at 4.6 eV. In overall, the absorption spectra Figure 6a-e identify several promising 2D semiconductors with optical properties similar or better than those of the well-known H-MoS$_2$. We also find that combining the materials in a vertical heterostructure improves the overall optical response.[43–45] The 5 materials (Figure 6(a-e)) with strongest response were stacked in a heterostructure and the absorbance was calculated using the transfer matrix method. The heterostructure shows larger *A* (Figure 6f) over a broadband spectrum, when compared with individual materials.

Inclusion of excitonic effects also modulates the reflectance of 2D materials. *R* increases at excitonic resonance in 2D materials as shown in Figure S3 (Supporting Information). Excitonic resonances in case of MoS$_2$ are denoted by A, B, and C peaks, which occur at ~1.9, ~2.1 and ~2.66 eV, respectively as shown in Figure 6b. This resonance between excitonic ground state and its first excited state is similar to the case of perfect nesting (Figure 1b) in our two-band model. The *R*, *A* in such a case depends only on the relaxation time $\tau$, and larger $\tau$ gives higher *R*, *A*. The increase in *R*, *A* of 2D MoS$_2$ at excitonic resonance is also seen in our BSE calculations at different $\tau$ values (Figure S9, Supporting Information). Therefore, cleaner samples at low temperatures will have smaller scattering rates (larger $\tau$), and hence exhibit improved *R*, *A*. Recently, a *R* of >80% was experimentally observed in 2D MoSe$_2$ at excitonic resonance due to larger $\tau$ at low temperatures, signifying that sample quality and temperature play a significant role in improving the optical response.[40] In addition to cleanliness, the flatness of samples also affects the sharpness of the *TAR* peaks.[46] Depending upon the substrate flatness, the 2D material can experience local distortions leading to loss of structural symmetry, causing additional scattering and inhomogeneous broadening. Our calculations are for strictly flat, free standing, pristine materials, and hence inhomogeneous broadening is not included in our model. Here we use relaxation time approximation, where all the other scattering mechanisms (electron-electron, electron-phonon and radiative) are combined into a single homogeneous broadening term. In experiments, the flatness can be assured by using atomically smooth non-interacting substrates such as *h*-BN.[46]

Another important characteristic of optical materials is skin/penetration depth, which measures how deep the incident electromagnetic radiation penetrates the material. Since the TMD monolayers exhibit promising absorbance values (smaller skin depth), we extend our transfer matrix formalism to the $G_0W_0$+BSE analysis to estimate the skin depth of the select TMD monolayers at peak frequencies marked in Figure 6a-e. In Table 2 we report the distance at which the transmittance $T$ falls to $1/e^2$ of its value at the surface.

**Table 2. Skin depth of TMD stacked monolayers at frequencies corresponding to peaks in absorbance in Figure 6a-e. The skin depth of bulk silver is given for comparison.**

| material | peak frequency (eV) | skin depth (nm) | |
|---|---|---|---|
| | | material | bulk Ag |
| T-PtTe$_2$ | 1.58 | 4.96 | 18 |
| H-MoS$_2$ | 2.66 | 3.0 | 22 |
| BP | 4.6 | 1.1 | 25 |
| H-TiS$_2$ | 3.95 | 1.96 | 25 |
| T-SnSe$_2$ | 3.68 | 3.2 | 25 |

Our estimate of skin depth of H-MoS$_2$ is in good agreement with experimental observations of ~ 5 nm at ~3 eV,[47] with discrepancy arising from the quality of samples (*i.e.* different scattering times), and difference in interlayer distance. Hence, our estimates can be used to gain a quantitative as well as qualitative measure of comparison among the compounds. We find that BP exhibits the smallest skin depth of ~1 nm (4.6 eV), in agreement with its highest predicted absorbance of ~43%. In comparison, it is seen from Table 2 that the skin depth of bulk Ag is approximately 5-10 times larger than that of TMDs and that of stacked 2D boron reaches 19 nm at 2 eV, reaffirming the exceptional absorption capacity of 2D TMDs and BP, with possible applications for radiation shielding.

## CONCLUSIONS

In conclusion, we carried out a computational survey of the optoelectronic properties of 2D materials, in particular, their transmittance, absorbance, and reflectance (*TAR*), with the goal of identifying materials with strongest optical response in different frequency ranges. Analytically, we gain microscopic insights into the origins of maximum response. Using single-band model, we find that 2D metals have high reflectance in THz regime; however, for maximum broadband $R$, high carrier concentration is required. Our results show that vertical heterostructures formed from stacking of 2D boron have broadband reflectance from IR to UV range and outperform doped graphene and bulk silver, with reflectance reaching ~99% for >100 layers. Phenomenologically, 2D materials have absolute absorbance limit of $A_{\text{lim}} = 1/2$. The analytical two-band model elucidates the role of band nesting to attain $A_{\text{lim}}$. We find from $G_0W_0$+BSE calculations that among all 2D semiconductors considered, T-PtTe$_2$, H-MoS$_2$, H-TiS$_2$, T-SnSe$_2$, and black phosphorus, have absorbances ≳30% in the near-IR, visible, near-UV, mid-UV, and deep-UV regions, respectively, mediated by band nesting and excitonic effects. Low scattering rates are important for achieving $A_{lim}$ in these materials. Stacking these materials in a vertical heterostructure further improves the overall optical response. A larger $A$ approaching the $A_{lim}$ value over a broadband spectrum is achieved in the van der Waals heterostructures. The higher absorbance of BP and TMD monolayers is also manifested in their skin depth, which is at least 5-10 times smaller than bulk silver. These materials with maximum response in different optical regimes are ideal for compact optoelectronics and understanding light-matter interactions.

## METHODS

We carried out first-principles density functional theory (DFT) calculations with dielectric response as implemented in the VASP[48] and GPAW[49] codes. The linear dielectric response was estimated for both the non-interacting Kohn-Sham system under random phase approximation (RPA), as well as with self-energy corrections to KS single-particle eigenvalues and accounting for electron-hole interactions using $G_0W_0$+BSE formalism. The initial screening of optical properties (under RPA with PBE functional) of two-dimensional materials was done by extracting the dielectric function $\varepsilon(\omega)$ for structures listed in the van der Waals heterostructure database in the Computational Materials Repository[50,51]. Note that, the database comprises of structural information of 2D nonmagnetic semiconducting transition metal dichalcogenides and oxides with theoretical negative heats of formation and is also referenced in the Inorganic Crystal Structure Database.[52] Recently a large number of them were experimentally synthesized and a generalized procedure to make the others was proposed.[36]

The dielectric response beyond RPA was calculated using single-shot $G_0W_0$ procedure together with solution of the Bethe-Salpeter equation implemented in VASP. This technique correctly accounts for electron-hole interaction necessary to obtain an accurate excitonic spectra. A vacuum of 18 Å was used along out of plane direction to reduce the interaction between the periodic images. Spin–orbit coupling was included in all the calculations. A Γ-centered grid of 24×24×1 were used to sample the Brillouin Zone (BZ). The wave functions were expanded in a plane wave basis with energy cut-off of 500 eV. The Kohn-Sham orbitals obtained using PBE functional were used as a starting guess for $G_0W_0$ calculation. A plane wave cut-off of 333 eV and a frequency grid of 225 points were used for calculating the response function in the $G_0W_0$ approach. A total of $6N$ bands, where $N$ is the total number of valence electrons in the material was used in the dielectric function and $G_0W_0$ calculation. BSE calculations were performed using 50 bands (20 in valence and 30 in conduction band). These parameters were optimized to obtain a converged optical spectrum (see S12, Supporting Information for convergence results).

## ASSOCIATED CONTENT

**Supporting Information**

The Supporting Information is available free of charge on the ACS Publications website at DOI:

Transmittance, absorbance, and reflectance (*TAR*) of a 2D material using transfer matrix method, Lindhard dielectric function for a 2D electron gas, maximum reflectance condition, JDOS and maximum absorbance condition, RPA vs. BSE dielectric information, *R* and *A* of 53 monolayers calculated with unconverged $G_0W_0$+BSE, comparison of *TAR* for monolayers of graphene and boron, comparison of the reflectance of heterostructures of 2D boron and doped graphene, dielectric response in the 2D limit, transmittance of 53 monolayers in the UV range, *R*, *A* of $MoS_2$ at different values of τ, convergence results, and validation of transfer matrix method.

## AUTHOR INFORMATION

**Corresponding Author**
*Email: biy@rice.edu

## ACKNOWLEDGMENTS

The authors acknowledge the DAVinCI - Center for Research Computing at Rice University, DoD HPCMP, and National Energy Research Scientific Computing Center (NERSC), a U.S. Department of Energy Office of Science User Facility for computational resources.